\documentstyle[11pt]{article}
\newcommand{\beq}{\begin{equation}}
\newcommand{\eeq}{\end{equation}}
\newcommand{\beqr}{\begin{eqnarray}}
\newcommand{\eeqr}{\end{eqnarray}}
\newcommand{\nn}{\nonumber}
\newcommand{\cosec}{{\rm cosec}}
\newcommand{\sech}{{\rm sech}}
\newcommand{\cosech}{{\rm cosech}}
\newcommand{\sss}{\vspace{.2in}}

\begin{document}
\begin{titlepage}
\vspace{-.2in}\begin{flushright}\today\end{flushright}
\begin{center} {\Large {\bf  
Shape Invariant Natanzon Potentials from Potential Algebra} }\\

\vspace*{1 in}

{\large Asim Gangopadhyaya$^{a,}$\footnote{e-mail: agangop@luc.edu, 
asim@uic.edu}, Jeffry V. Mallow $^{a,}$\footnote{e-mail: jmallow@luc.edu}
and Uday P. Sukhatme$^{b,}$\footnote{e-mail: sukhatme@uic.edu}.}
\end{center}
\begin{tabular}{l l}
$a$) &Department of Physics, Loyola University Chicago,
Chicago, IL 60626;\\
$b$) &Department of Physics, University of Illinois at
Chicago, (m/c 273) \\& 845 W. Taylor Street, Chicago, IL 60607-7059.\\
\end{tabular}
\vspace{1.0in}
\begin{center}
{\large {\bf Abstract}}\\
\end{center}

Using the underlying algebraic structures of Natanzon potentials, we discuss 
conditions that
generate shape invariant potentials. In fact, these conditions give all the 
known shape invariant
potentials corresponding to a translational change of parameters. We also 
find that while the
algebra for the general Natanzon potential is $SO(2,2)$, a subgroup $SO(2,1)$ 
suffices for all the
shape invariant problems of Natanzon type.
\end{titlepage}
%\end{document}
\newpage
%****************************************
\noindent {\large {\bf I. Introduction}}
%****************************************

\sss
Natanzon potentials\cite{Natanzon} form a complete set of exactly solvable 
potentials of nonrelativistic quantum mechanics for which the 
Schr\"odinger equation reduces to the hypergeometric equation. 
In refs. \cite{Alhassid,Wu}, Alhassid et al. have studied the 
group structure of these quantum mechanical systems,  related their
Hamiltonians to the Casimir operator of an underlying SO(2,2) algebra, and 
determined 
all their quantum states by group theoretical methods.

In supersymmetric quantum mechanics (SUSY-QM)\cite{Cooper}
%.\cite{Witten,Cooper_Freedman}, 
one applies a different algebraic method. The exactly solvable 
problems in SUSY-QM are described by 
superpotentials $W(x,a)$ that obey a special integrability condition,
\beq
W^2(x,a_0)+\frac{d~W(x,a_0)}{d~x}=
W^2(x,a_1)-\frac{d~W(x,a_1)}{d~x}+R(a_0) ~,
\label{SIC}
\eeq
known as shape invariance\cite{Infeld,Gendenshtein}. $R(a_0)$ 
is a constant and the parameter $a_1$ is a function of $a_0$, i.e. 
$a_1=f(a_0)$. For a shape invariant system, the entire spectrum can be 
determined algebraically
by a procedure similar to that of the one dimensional harmonic oscillator, 
without ever referring
to the underlying differential equations. 
Although most of the known shape invariant potentials (SIP) belong 
to the Natanzon class, there
are a few exceptions \cite{Spiridinov,Barclay}.

In a previous work\cite{Gangopadhyaya_proc}, we have shown that 
problems for which 1) there
is a translational change of parameters $a_1 = f(a_0) = a_0+~{\rm constant}$~ 
\underline{and}~
2) $R(a_0)$ is a linear function of $a_0$, the shape invariance condition 
of eq. (\ref{SIC})
implies the presence of a $SO(2,1)$ dynamical algebra. Hence these 
problems are solvable by
either method. As shown in ref. \cite{Cooper}, potentials of this type includes 
Morse, Scarf I,
Scarf II, and generalized P\"oschl-Teller. However, the formalism used in ref.
\cite{Gangopadhyaya_proc} is not readily extendable to other shape 
invariant potentials.

In this paper, we generalize our work of ref. \cite{Gangopadhyaya_proc}. 
However, we have used a different approach here which is 
closely based on the work of Alhassid
et. al. \cite{Alhassid}. The authors of ref. \cite{Alhassid} have shown 
that a Hamiltonian with a
general Natanzon potential has a $SO(2,2)$ symmetry. We study here  the 
algebra of Natanzon
potentials that are also shape invariant. We find that general Natanzon  
potentials when subjected
to a further constraint give the entire set of shape invariant potentials.
The shape invariant potentials that reduce to the confluent
hypergeometric equation can be obtained as a limit\cite{Gangopadhyaya_pct}.  We 
also find that
while the algebra for the general Natanzon potential is $SO(2,2)$, a subgroup 
$SO(2,1)$ suffices
for all the shape invariant problems of Natanzon type.

Thus, this paper connects all the shape invariant potentials of translational 
type $(a_1=a_0+{\rm
constant})$ to an algebraic structure which has many interesting consequences. 
Some time ago it
was discovered that spectra of potentials with translational shape invariance 
can be exactly
determined by the supersymmetric WKB method\cite{Comtet}, which usually only 
gives
approximate results. The reason for this exactness was very puzzling. However, 
in light of this
group theoretical connection, this result may not be that difficult to 
understand as various authors
have demonstrated the exactness of WKB results on a group manifold 
\cite{Funahashi}. 

In sec. II, we will quickly review the formalism of SUSY-QM and shape 
invariance. In sec. III, we
will briefly describe our previous work \cite{Gangopadhyaya_proc} where we 
connected a subset
of SIPs to $SO(2,1)$ potential algebra. In sec. IV, we discuss the potential 
algebra of a general
Natanzon potential. This section will closely follow ref. \cite{Alhassid}. In 
sec. V, we identify conditions under which the general Natanzon potential 
reduces to a shape invariant potential. We
then show that this condition has a finite number of solutions for shape 
invariant potentials;
however, they generate all the known shape invariant potentials of 
translational type. 
\\

\sss

%**************************************************************
\noindent {\large {\bf II. SUSY-QM and Shape Invariance}}
%**************************************************************

\sss 
In this section, we very briefly describe supersymmetric quantum 
mechanics (SUSY-QM), 
and also show how SUSY-QM with shape invariance allows one to
completely determine the spectrum of a quantum system. For a more detailed
description, see ref. \cite{Cooper}.

A quantum mechanical system given by a potential $V_{-}(x)$ can 
alternatively be described by its ground state wavefunction 
$\psi^{(-)}_0$. From the Schr\"odinger equation for the ground state 
wavefunction 
$\left(  - \psi^{''}_0 + V(x) \psi_0 =0\right)$, 
it follows that the potential can be written as, 
$V_{-}(x)=\left({{\psi^{''}_0} \over {\psi_0}}\right)$, where prime denotes 
differentiation with respect to  $x$.  [It is assumed that the 
potential is properly adjusted to make the ground state energy $E_0=0$].  
In SUSY-QM, it is customary to express the system in terms of the 
superpotential $W(x)=-\left({{\psi^{'}_0} \over {\psi_0}}\right)$. The
ground state wavefunction is then given by 
${\psi_0} \sim \exp\left( - \int^x_{x_0} W(x) dx \right)$; $x_0$ is an 
arbitrarily chosen reference point. 
The Hamiltonian $H_{-}$ can now be written as  
\begin{eqnarray}
H_{-} 
&=& \left(-{{d^2} \over {dx^2}}+ V_{-}(x) \right)\;=\;
\left(-{{d^2} \over {dx^2}}+ W^2(x)-{{dW(x)} \over {dx}}\right) 
\\
&=& \left(-{d \over {dx}} +W(x)\right)  
\left({d \over {dx}} +W (x) \right) 
\nn 
\label{schrodinger}
\end{eqnarray}
(We are using units with $\hbar$ and $2m=1$.)
In analogy with the harmonic oscillator raising and lowering operators, we 
have introduced two operators $A = \left({d \over {dx}} +W(x)\right)$, and
and its Hermitian conjugate $A^{+}=\left(-{d \over {dx}} +W(x)\right)$. Thus
$H_-=A^\dagger   A$.

However, one can easily construct another Hermitian operator 
$H_+=A   A^\dagger$ and show that the eigenstates of $H_+$ are 
iso-spectral with 
excited states of $H_-$. The Hamiltonian $H_+$, with potential 
$V_{+}(x)=\left(W^2(x)+{{dW(x)} \over {dx}}\right)$, is called the 
superpartner of the Hamiltonian $H_-$. To show the iso-spectrality mentioned 
above, let us denote the eigenfunctions of $H_{\pm}$ that 
correspond to eigenvalues $E^{\pm}_n$, by $\psi^{(\pm)}_n$. 
For  $n=1,2, \cdots ~$, 
\begin{equation}
H_{+} \left( A \psi^{(-)}_n \right) = AA^{+} \left( A \psi^{(-)}_n
\right)
=A \left( A^{+}A \psi^{(-)}_n \right) =A H_{-} \left( \psi^{(-)}_n
\right)
=E^{-}_n \left( A \psi^{(-)}_n \right)\;.
\end{equation}
Hence, except for the ground state which obeys $A\psi^{(-)}_0=0$, all other 
states $\psi^{(-)}_n$ of $H_{-}$ there exists a state 
$\psi^{(+)}_{n-1} \propto A \psi^{(-)}_n$
of $H_+$ with exactly the same energy, i.e. $E^{+}_{n}\;=\;E^{-}_{n+1}$, 
where $n=0,1,2, \cdots $. Conversely, one also has 
$A^+ \psi^{(+)}_n \propto \psi^{(-)}_{n+1}$.

Thus, if the eigenvalues and the eigenfunctions of $H_{-}$ were
known, one would automatically obtain the eigenvalues and the
eigenfunctions of $H_{+}$, which is in general a completely different 
Hamiltonian.

Now, let us consider the special case where $V_-(x,a_0)$ is a shape invariant 
potential.  For such
systems, potentials $V_+(x,a_0)=V_-(x,a_1)+R(a_0)$. Their superpotential $W$ 
obeys the
integrability condition of eq. (\ref{SIC}). Since potentials $V_{+}(x,a_0)$ 
and $V_{-}(x,a_1)$
differ only in an additive constant, their common ground state wavefunction is 
given by
${\psi^{(-)}_0(x,a_1)} \sim \exp\left( - \int^x_{x_0} W(x,a_1) dx \right).$ 
The ground state
energy of $H_{+}(x,a_0)$ is $R(a_0)$, because the ground state energy of 
$V_{-}(x,a_1)$
vanishes. Now using SUSY-QM algebra, the first excited state of $H_-(x,a_0)$ is 
given by
$A^+{(x,a_0)} \psi^{(-)}_0(x,a_1)$ and the corresponding eigenvalue is 
$R(a_0)$. By itereating
this procedure, the $(n+1)$-th excited state is given by 
$$\psi^{(-)}_{n+1}(x,a_0) \sim
A^+{(a_0)} ~ A^+{(a_1)}  \cdots  A^+{(a_n)} ~ \psi^{(-)}_0(x,a_n)\;,$$ and 
corresponding
eigenvalues are given by $$E_0=0; 
~~{\rm and}~~E_n^{(-)}=\sum_{k=0}^{n-1}R(a_k)~~{\rm
for}~n>0.$$ (To avoid notational complexity, we have suppressed the 
$x$-dependence of
operators $A(x,a_0)$ and $A^{+}(x,a_0)$.) Thus, for a shape invariant 
potential, one can obtain
the entire spectrum of $H_-$ by the algebraic methods of SUSY-QM.

\sss

Most of the known exactly solvable problems possess a spectrum generating 
algebra
(SGA)\cite{Alhassid,Wu,Barut}. We would 
like to find out if there is any connection between SGA and shape invariance 
of these systems.
As we shall see later, the type of SGA that is most relevant to us is known as 
potential algebra,
studied extensively by Alhassid et al.\cite{Alhassid,Wu}. In potential algebra, 
the Hamiltonian of
the system is written in terms of the Casimir operator ($C_2$) of the algebra, 
and the energy of
states specified by an eigenvalue $\omega$ of $C_2$ is fixed. Different states 
with fixed
$\omega$ represent eigenstates of a set of Hamiltonians that differ only in 
values of parameters
and share a common energy. For a system with a $SO(2,1)$ potential algebra, 
the different
values of parameters are eigenvalues of operator $J_3$, chosen to form a 
complete set of
commuting observables. This is very similar to the case of shape invariant 
potentials. In the next
section, we will attempt to establish this 
connection in a more concrete fashion. In fact, for a set of solvable 
quantum mechanical systems we shall explicitly show that shape invariance 
leads to a potential
algebra. 

\sss

%%%%%%%%%%%%%%%%%%%%%%%%%%%%%%%%%%%%%%%%%%%%%%
\noindent {\large {\bf III. Shape Invariance and Connection to Algebra}}
%%%%%%%%%%%%%%%%%%%%%%%%%%%%%%%%%%%%%%%%%%%%%%
\sss

Let us consider a generic shape invariant potential $V_-(x,a_0)$ with a
translational change of parameters $a_{m+1}=a_{m}+\delta=a_0+(m+1)\delta$,
where $\delta$ is a constant. For the superpotential 
$W \left(x,a_m \right)\equiv W(x,m)$, the
shape invariance condition is
\begin{equation}
W^2(x,m)+W'(x,m)=W^2(x,m+1)-W'(x,m+1)+R(m)~~.
\label{SIC2}
\end{equation}
It is natural to ask whether the change of parameters can be formally 
accomplished by the action of a ladder type operator. 
With this in mind, we define an operator 
$J_3= - i\frac{\partial}{\partial \phi} \equiv -i\partial_\phi$, analogous 
to the $z$-component of
the angular momentum operator. It acts upon functions in the space described 
by two coordinates
$x \;{\rm and}\; \phi$, and its eigenvalues $m$ will play the role of the 
parameter of the
potential. We also define two more operators, $J^-$ and its Hermitian 
conjugate $J^+$ by
\begin{equation}
J^{\pm} = 
e^{\pm i\,\phi} 
\left[ \pm \frac{\partial}{\partial x} -
W \left(x, -i\,\partial_\phi 
\pm  
\frac{1}{2}
\right) \right]  \;\;\;
\;.
\label{J_pm}
\end{equation}
The factors $e^{\pm i\,\phi}$ in $J^{\pm}$ ensure that they indeed behave
as ladder operators for the quantum number $m$. Operators $J^{\pm}$ are 
basically of the same form as the $A^\pm$ operators of SUSY-QM, except that 
the parameter $m$ of the superpotential $W$ is replaced by operators 
$\left( J_3 \pm  \frac{1}{2} \right)$.
Explicit computation shows that 
\begin{equation}
\left[J_3,J^\pm \right] = \pm J^\pm ~~,
\end{equation}
and hence operators $J^\pm$  change the eigenvalues of the $J_3$ operator by 
unity, similar to the ladder operators of angular momentum ($SO(3)$). 
Now let us determine the commutator $\left[J^+,J^- \right] $.
\begin{eqnarray}
\label{Jp,Jn}
\left[J^+,J^-\right] 
&=& J^+J^- - J^-J^+=\left[ -\frac{\partial^2}{\partial x^2}+ 
W^2 \left( x,J_3-\frac{1}{2} \right) -
W' \left( x,J_3-\frac{1}{2} \right) 
\right]\; 
\nonumber \\
& &\;-\;
\left[ -\frac{\partial^2}{\partial x^2}+ 
W^2 \left( x,J_3+\frac{1}{2} \right) +
W' \left( x,J_3+\frac{1}{2} \right) 
\right]\; 
\nonumber \\
&=& -R \left( J_3+\frac{1}{2} \right)\;, 
\end{eqnarray}
where we have used the shape invariance condition (\ref{SIC2}).
Thus, we see that shape invariance enables us to close the algebra 
of $J_3$ and $J^\pm$ to 
\begin{equation}
\left[J_3,J^\pm \right] = \pm J^\pm~~,~~~~
\left[J^+,J^-\right] = -R \left( J_3+\frac{1}{2} \right)~~.
\label{algebra}
\end{equation}

Now, if the function $R(J_3)$ is linear in $J_3$, the algebra of 
eq. (\ref{algebra}) reduces to that of $SO(3)$ or $SO(2,1)$ 
\cite{Gangopadhyaya_proc}. 
Several SIP's are of this type,  among them are the Morse, Scarf~ I, Scarf ~II, 
and generalized P\"oschl-Teller potentials. For these potentials, 
$R\left( J_3+\frac{1}{2}
\right)=2~J_3$\cite{Cooper}, and eq. (\ref{algebra}) reduces to an 
$SO(2,1)$ algebra and hence establishes 
a connection between shape invariance and potential algebra. 
With a slightly different formalism, Balantekin arrived at a similar conclusion 
for
these SIPs\cite{Balantekin}.  However, there are many other important systems 
like Coulomb,
Eckart etc. where $R(a_0)$ is not linear in $a_0$, and these cases will be 
discussed later.

\sss

%%%%%%%%%%%%%%%%%%%%%%%%%%%%%%%%%%%%%%%%%%%%%%
\noindent {\large {\bf IV. Differential Realization of SO(2,2)}}
%%%%%%%%%%%%%%%%%%%%%%%%%%%%%%%%%%%%%%%%%%%%%%

\sss

Before establishing a connection between a general Natanzon Hamiltonian and
a $SO(2,2)$ potential algebra, we will discuss a realization of $SO(2,2)$ 
algebra in terms of differential operators on a $(2,2)$-hyperboloid. 
For consistency, we use the formalism and the notations of 
refs. \cite{Alhassid,Wu}.

A $(2,2)$-hyperboloid is defined by 
\beqr
x_1=\rho ~\cosh \chi ~\cos \phi, && x_2=\rho ~\cosh \chi ~\sin \phi \nn \\
x_3=\rho ~\sinh \chi ~\cos \theta, && x_4=\rho ~\sinh \chi ~\sin \theta 
~,
\label{transform}
\eeqr
where $\phi$ and $\theta$ are rotation angles in the $x_1,x_2$ and 
$x_3,x_4$ planes respectively $[0 \leq \phi,\theta <2\pi]$. Six generators of 
the algebra, $J_i$ and $K_i$ ($i=1, \cdots ,3$) can be chosen as
\beqr
J_1=(x_2~p_3+x_3~p_2)~, & J_2=-(x_1~p_3+x_3~p_1)~, & J_3=(x_1~p_2-x_2~p_1)
\nn \\
K_1=(x_1~p_4+x_4~p_1)~, & K_2=(x_2~p_4+x_4~p_2)~, & K_3=(x_3~p_4-x_4~p_3 )~.
\eeqr
Operators $p_i$ represent derivatives $-i\frac{\partial}{\partial x_i}$. 
The algebraic relations obeyed by these operators are given by
\beqr
[J_1,J_2] = -iJ_3, & [J_2,J_3] = iJ_1, & [J_3,J_1] = iJ_2, \nn 
\eeqr
\beqr
[K_1,K_2] = -iJ_3, & [K_2,K_3] = iJ_1, & [K_3,K_1] = iJ_2, \nn 
\eeqr
\beqr
[J_1,K_2] = -iK_3, & [J_2,K_3] = iK_1, & [J_3,K_1] = iK_2, \nn 
\eeqr
\beqr
[K_1,J_2] = -iK_3, & [K_2,J_3] = iK_1, & [K_3,J_1] = iK_2. 
\label{SO(2,2)_algebra}
\eeqr
The above algebra can be decomposed in terms of two 
commuting $SO(2,1)$ algebras generated by 
\beq
A_i={1 \over 2} \left( J_i+K_i\right); ~~~~
B_i={1 \over 2} \left( J_i-K_i\right)~.
\eeq
These operators commute, i.e. $[A_i,B_j] = 0$. 
Using eqs. (\ref{transform}) and (\ref{SO(2,2)_algebra}), the differential 
realization can be written explicitly as \cite{Alhassid,Wu}
\beqr
A^{\pm} \equiv A_1 \pm A_2  &=& {1\over 2}~e^{\pm i(\phi+\theta)}
     \left[ \mp \frac{\partial}{\partial \chi} + 
     \tanh \chi \left(-i \partial_\phi \right) +
     \coth \chi \left(-i \partial_\theta \right) \right];
\\
A_3&=& -{i \over 2} \left(\partial_\phi + \partial_\theta \right); 
\nn \\
B^{\pm} \equiv B_1 \pm B_2  &=& {1\over 2}~e^{\pm i(\phi-\theta)}
     \left[ \mp \frac{\partial}{\partial \chi} + 
     \tanh \chi \left(-i \partial_\phi \right) +
     \coth \chi \left(+i \partial_\theta \right) \right];
\nn \\
B_3&=& -{i \over 2} \left(\partial_\phi - \partial_\theta \right)~ .
\nn
\label{Apm}
\eeqr
The $SO(2,1)$ algebra obeyed by these operators is 
\beqr
[A_3, A^\pm] = \pm A^\pm, && [A^+,A^-] = -2A_3 \nn 
\eeqr
and a similar one for the $B$'s. The Casimir operator $C_2$ is given by 
\beqr
C_2 &=& 2~\left(A_3^2 - A_+A_- - A_3 \right) 
     + 2~\left(B_3^2 - B_+B_- - B_3 \right) \nonumber \\
    &=&
     \left[ \frac{\partial^2}{\partial \chi^2} 
     +\left( \tanh \chi + \coth \chi \right) 
      \frac{\partial}{\partial \chi} 
     + \sech^2\chi    \left(-i \partial_\phi \right)^2
     - \cosech^2 \chi \left(-i \partial_\theta \right)^2
     \right]  .
\label{casimir}
\eeqr
Operators $A_3$, $B_3$ and $C_2$ can be simultaneously diagonalized, and their
actions on their common eigenstate are given by
\beqr
C_2 |\omega,m_1,m_2 \rangle &=& \omega (\omega+2) ~|\omega,m_1,m_2 \rangle ~;
\nn \\
A_3 |\omega,m_1,m_2 \rangle &=& m_1 ~|\omega,m_1,m_2 \rangle ~;
\nn \\
B_3 |\omega,m_1,m_2 \rangle &=& m_2 ~|\omega,m_1,m_2 \rangle~.
\label{action}
\eeqr

It is worth mentioning at this point that the Casimir operator given above 
is indeed self-adjoint with respect to a measure 
$\sinh\chi\cosh\chi d\chi d\phi d\theta$.

\sss

%%%%%%%%%%%%%%%%%%%%%%%%%%%%%%%%%%%%%%%%%%%%%%
\noindent {\large {\bf V. The Natanzon Potentials}}
%%%%%%%%%%%%%%%%%%%%%%%%%%%%%%%%%%%%%%%%%%%%%%

\sss
The Schrodinger equation for any Natanzon potential can be reduced by a point 
canonical transformation (a general similarity transformation followed by an 
appropriate change of independent variable)
\cite{Gangopadhyaya_pct,Cooper1,De} 
to the hypergeometric equation. A general potential $U(r)$ of the 
Natanzon type is implicitly defined by \cite{Natanzon} 
\beq
U[z(r)] = \frac{-f   z(1-z) + h_0   (1-z) + h_1   z}{R(z)}
-{1\over 2} \left\{ z,r \right\}~,
\label{Natanzon_pot}
\eeq
where $R(z)=a z^2 +b_0 z + c_0 = a (1-z)^2 -b_1 (1-z)+ c_1$ and 
$f, h_0, h_1, a, b_0, b_1, c_0, c_1$ are constants. The Schwarzian derivative
$\left\{ z,r \right\}$ is defined by 
\beq
\left\{ z,r \right\} \equiv \frac{d^3z/dr^3}{dz/dr} -{3 \over 2}
\left[ \frac{d^2z/dr^2}{dz/dr} \right]^2 ~.
\label{Schwarzian}
\eeq
The relationship between variables $z$ ($0<z<1$) and $r$ is implicitly given by
\beq
\left( \frac{dz}{dr} \right) = \frac{2z(1-z)}{\sqrt{R(z)}}~.
\label{z,r}
\eeq
To avoid a singularity in $U\left( z(r) \right)$, one assumes that $R(z)$ has 
no singularity in the
domain $(0,1)$. The Schr\"odinger equation is given by 
\beqr
  \left[ \frac{d^2}{dr^2} + \left\{
     \left( \frac{dz}{dr} \right)^2 I(z) +  
     {1\over 2} \left\{ z,r \right\}
     \right\}
  \right]~=~0, &&\nn 
\eeqr
where 
\beqr
I(z) &=& \frac{\left( 1-\lambda_0^2  \right) \left( 1-z \right) + 
\left( 1-\lambda_1^2  \right) z + \left( \mu^2 - 1 \right) z \left( 1-z 
\right)}{4z^2\left( 1-z \right)^2} \nn 
\eeqr
and
\beqr
\left( 1-\mu^2  \right) = a~E-f, 
& \left( 1-\lambda_0^2  \right) = c_0~E-h_0, 
& \left( 1-\lambda_1^2  \right) = c_1~E-h_1 ~.
\eeqr
To connect the Casimir operator $C_2$ of the $SO(2,2)$ algebra [eq. 
(\ref{casimir})] to the
general Natanzon potential, we will first perform a similarity transformation 
on $C_2$ by a
function $F$ and then follow that up by an appropriate change of variable 
$\chi \rightarrow g(r)$.
Under the similarity transformation, 
$$\frac{d}{d\chi} \longrightarrow F   \frac{d}{d\chi}   F^{-1} =
\left(\frac{d}{d\chi} - \frac{{\dot F}}{F} \right)~~,
~~~~\frac{d^2}{d\chi^2} \longrightarrow  \left(\frac{d^2}{d\chi^2} 
- \frac{2{\dot F}}{F} \frac{d}{d\chi} +\frac{2{\dot F}^2}{F^2} 
- \frac{{\ddot F}}{F}  \right),$$
where dots represent derivatives with respect to $\chi$.  The Casimir operator 
$C_2$ of eq. (\ref{casimir}) transforms as:
\beqr
C_2 \longrightarrow {\tilde C}_2
& = & \left[ \frac{d^2}{d\chi^2} 
 +      \left(\tanh\chi+\coth\chi - \frac{2 {\dot F} }{F} \right) 
                    \frac{d}{d\chi} 
        + \frac{2 {\dot F}^2}{F^2} 
     -  \frac{{\ddot F}}{F}  
      \right. \nn \\ 
&&    \left. 
     - \left(\tanh\chi+\coth\chi \right) \frac{{\dot F} }{F} 
     + \sech^2\chi \left(-i\partial_\phi \right)^2
     - \cosech^2\chi \left(-i\partial_\theta \right)^2
      \right]  ~~.
\eeqr
Now, let us carry out a change of variable from $\chi$ to $r$ via $\chi=g(r)$.
We are going to denote differentiation with respect to $r$ by a prime.
Operators $\frac{d}{d\chi}$ and $\frac{d^2}{d\chi^2}$ transform as 
$$\frac{d}{d\chi} = \frac{1}{g'}\frac{d}{dr} ~~,
~~~~~~~~~~ \frac{d^2}{d\chi^2} = \frac{1}{g'^2}\left[ \frac{d^2}{dr^2}
-\frac{g''}{g'}  \frac{d}{d r} \right] ~.$$
The operator 
${\tilde C}_2$ now transforms into 
\beqr
{\tilde C}_2 &=& \frac{1}{g'^2} 
\left[ \frac{d^2}{dr^2}
+ \left\{ -\frac{g''}{g'} - \frac{2F''}{F} 
+ g'   \left(\tanh  g +\coth  g  \right) \right\} \frac{d}{d r} 
\right.   \nn\\  &&  \left. 
+ \frac{2 F'^2}{F^2} - \frac{F''}{F} +\frac{F'~g''}{F~g'} 
      \right. \nn \\ 
&&    \left. 
- \frac{F'~g'}{F} \left( \tanh  g +\coth  g  \right) 
+ g'^2 \left(   \sech^2  g  \left(-i \partial_\phi \right)^2
-           \cosech^2  g  \left(-i \partial_\theta\right)^2 
       \right)
      \right]~~. 
\label{tilde_C_2}
\eeqr

\sss
In order for $g'^2 {\tilde C}_2 $ to be a Schr\"odinger Hamiltonian, we require
the expression inside the curly brackets in eq. (\ref{tilde_C_2}) to vanish.
This constrains the relationship between the two functions $F$ and $g$ to be 
\beq
-\frac{g''}{g'} - \frac{2 F'}{F} 
+ g'   \left(\tanh g +\coth g  \right) =0,
\eeq
which yields 
\beq
F \sim  \left( \frac{\sinh (2g)}{g'}\right)^{1\over 2}~.
\eeq
Thus, the operator $\tilde{C}_2$, transforms into 
\beqr
{\tilde C}_2 &=& \frac{1}{g'^2} 
\left[ \frac{d^2}{dr^2}
+ g'^2 \left( \frac{\left( 1-\tanh^2g \right)^2 - 4\tanh^2g }{4 
        \tanh^2g}\right) 
      \right. \nn \\ 
&&    \left. 
+{1\over 2} \left\{g,r\right\} 
+ g'^2 \left( \sech^2 g \left(-i\partial_\phi \right)^2
     - \cosech^2 g \left(-i\partial_\theta \right)^2
       \right)
      \right]~~.
\eeqr
This Casimir operator now has a form of $${\tilde C}_2 =- \frac{1}{g'^2} H,$$
where $H$ is a one-dimensional Hamiltonian with the potential $U(r)$ given by 
$$E-U(r)=
 g'^2 \left( \frac{\left( 1-\tanh^2g \right)^2 - 4\tanh^2g }{4 \tanh^2g}\right) 
+{1\over 2} \left\{g,r\right\} 
+ g'^2 \left[ \sech^2 g \left(-i\partial_\phi \right)^2
             -\cosech^2g \left(-i\partial_\theta \right)^2
       \right]~~.$$
Now, for this potential to take the form of a general Natanzon potential,
we have to relate variables $g$ and $z$ in such a way that the potential in 
terms of $z$ is given by eq. (\ref{Natanzon_pot}). Since the potential has to 
be a ratio of two quadratic functions of $z$, this is accomplished with the 
identification $z=\tanh^2g$, which leads to
\beqr
U(z(r))&=&\frac{E~R+[-{7 \over 4}+{5 \over 2}z-
{7 \over 4} z^2]-z(1-z) \left(-i\partial_\phi \right)^2
+(1-z)\left(-i\partial_\theta \right)^2 }{R} -{1\over 2} \left\{z,r\right\} 
\nn\\
&&\nn\\
&=&\left[
-\left( aE-{7 \over 4}+\left(-i\partial_\phi \right)^2 \right)z(1-z) 
+ \left(c_0 E -{7 \over 4} + \left(-i\partial_\theta \right)^2 \right) (1-z) 
\right. \nn \\
&& \left. +\left( (a+b_0+c_0) E -1 \right) \right] / R(z) 
~-{1\over 2} \left\{z,r\right\} 
~~.
\label{Natanzon2}
\eeqr
We have used 
$$g'=\frac{dg}{dz}~z'= \frac{1}{2 \sqrt{z} (1-z)} ~ 
\frac{2z(1-z)}{R}= \sqrt{\frac{z}{R}}~~.$$
Now, with the following identification
\beqr
f   &=& aE-{7 \over 4}+\left(-i\partial_\phi \right)^2 , \nn\\
h_0 &=& c_0 E -{7 \over 4} + \left(-i\partial_\theta \right)^2 ,\nn\\
h_1 &=& (a+b_0+c_0)E  -1 ~~,
\eeqr
the potential of eq. (\ref{Natanzon2}) indeed has the form of a general 
Natanzon potential [eq.
(\ref{Natanzon_pot})].
\sss

%******************************************************
\noindent {\large {\bf VI. Shape Invariant  Natanzon Potentials from 
Potential Algebra}}
%******************************************************

\sss
At this point we go back to the operators $A^\pm$ [eq. (\ref{Apm})] and see 
how they transform under the similarity transformation given by 
$F \sim \left( \frac{\sinh (2g)}{g'}\right)^{1\over 2} \sim \sqrt{z\over{z'}}$. 
This transformation carries operators $A^\pm$ to 
\beq
A^\pm \longrightarrow {\tilde A}^\pm = 
     {{e^{\pm i (\phi+\theta)} }\over 2} 
\left[
     \mp \left( \frac{d}{d\chi}  + \frac{1}{2z'} \frac{dz'}{d\chi}
           - \frac{1}{2z} \frac{dz}{d\chi} \right) 
           +\tanh\chi ~(-i\partial_\phi)+\coth\chi ~(-i\partial_\theta)
\right]~.
\eeq
If the expression $\left(\frac{1}{2z'} \frac{dz'}{d\chi}-\frac{1}{2z} 
\frac{dz}{d\chi} \right)$ can
be written as a linear combination of $\tanh\chi$ and $\coth\chi$, operators 
${\tilde A}^\pm$ can
be cast in a form similar to the operators $J^\pm$ of eq. (\ref{J_pm}), and 
connection with shape invariance is established. 

Hence to get shape invariant potentials, we require, $$\left( \frac{1}{2z'}
\frac{dz'}{d\chi}-\frac{1}{2z} \frac{dz}{d\chi} \right) =\alpha\tanh\chi + 
\beta\coth\chi.$$ This leads to 
$z'=z^{1+\beta} \dot (1-z)^{-\alpha-\beta}$, which is another restriction 
on the relationship between variables $z$ and $r$. Since these variables are 
already constrained by eq. (\ref{z,r}), only a handful of solutions would be 
compatible with both restrictions. Thus $z(r)$'s that are compatible with both 
equations are given by
\beq
z^{1+\beta} \dot (1-z)^{-\alpha-\beta}=\frac{2z(1-z)}{\sqrt{R(z)}},
\label{constraint}
\eeq
where $R(z)$ is a quadratic function of $z$. After some computation, we find 
that there are only a finite number of values of $\alpha$, $\beta$ which 
satisfy eq. (\ref{constraint}).
These values are listed in Table 1, and they exhaust all known shape invariant 
potentials that lead
to the hypergeometric equation. It is also interesting to note that while the 
potential algebra of a
general Natanzon system is $SO(2,2)$, and requires two sets of raising and 
lowering operators
$A^{\pm}$ and $B^{\pm}$, all known shape invariant potentials need only one 
such set. For all
SIPs of Table 4.1 of ref. \cite{Cooper}, one finds that all partner potentials 
are connected by
change of just one independent parameter (although other parameters which don't 
change are also
present.) Thus there is a series of potentials that only differ in one 
parameter. 
From the potential
algebra perspective, all these potentials  differ only by the eigenvalue of an 
operator that is a
linear combination of $A_3$ and $B_3$, and all are characterized by a common 
eigenvalue of
$C_2$. Thus, these shape invariant potentials can be associated with a 
$SO(2,1)$ potential
algebra generated by operators $A^+$, $A^-$ and the same linear combination of 
$A_3$ and $B_3$. \\

%******************************************************
%                ACKNOWLEDGEMENT
%******************************************************

A.G. acknowledges a research leave from Loyola University Chicago which
made his involvement in this work possible. We would like to thank Prof. 
Y. Alhassid for clarifying several points. One of us (AG) would also
like to thank the Physics Department of the University of Illinois for 
warm hospitality. Partial financial support from the U.S. Department of 
Energy is gratefully acknowledged.
\newpage
%-----------------------------------------------------------------------

\vspace*{1.5in}

\begin{center}
\begin{tabular}{||c|c|l|l|l||}
\hline&&&&\\
{\bf $\alpha$} & {\bf $\beta$}  &  {\bf $z(r)$} &    {\bf Superpotential} &    
{\bf Potential}\\
&&&&\\
\hline&&&&\\
 $0$      &  $0$    & $z=e^{-r}$   &$ \tilde{m}_1 \coth \frac{r}{2} + 
\tilde{m}_2$      & Eckart\\
     &         &              &                        &\\
\hline&&&&\\
 $0$      &$ -{1\over 2}$     & $z=\sin^2\frac{r}{2}$  &$ \tilde{m}_1 \cosec 
\,r +  \tilde{m}_2 
\cot r $
                                        &        Gen.  P\"oschl-Teller\\
     &         &              &                   &   trigonometric\\
     &         &              &                        &\\
\hline&&&&\\
 $0$      & $-1$         &$z=1-e^{-r}$            &$ \tilde{m}_1 \coth 
\frac{r}{2} + \tilde{m}_2$ 
  &      Eckart \\
     &         &              &                        &\\
\hline&&&&\\
$-{1\over 2}$ &  $0$     &$z=\sech^2\frac{r}{2}$  &$ \tilde{m}_1 \cosech 
\,r +  \tilde{m}_2 
\coth r $
 & P\"oschl-Teller II \\
     &         &              &                        &\\
\hline&&&&\\
$-{1\over 2}$  & $-{1\over 2} $ &$z=\tanh^2\frac{r}{2} $&$ \tilde{m}_1 \tanh 
\frac{r}{2} 
+  \tilde{m}_2 \coth \frac{r}{2} $     &    Gen. P\"oschl-Teller \\
     &         &              &                        &\\
\hline&&&&\\
$-1$      &  $0$         &$z=1+\tanh \frac{r}{2}$      & $ \tilde{m}_1 \tanh 
\frac{r}{2} + \tilde{m}_2$
&        Rosen ~Morse\\
     &         &              &                        &\\
\hline
\end{tabular}

\vspace*{1.0in}
{Table 1}
\end{center}

\newpage
\noindent 
{\bf Table Caption:} Table 1 shows all allowed value of $\alpha,~\beta$ 
and the superpotentials that they generate. Constants $\tilde{m}_i,~(i=1,2)$ 
are linear functions of $m_1$ and $m_2$s of eq. (\ref{action}).
%-----------------------------------------------------------------------
\end{document}